\documentclass{article}
\usepackage{spconf,amsmath,graphicx}
\usepackage{subcaption}
\usepackage{amsfonts}
\usepackage{multirow}

\def\R{{\mathbb{R}}}
\def\softmax{{\mathrm{Softmax}}}
\def\att{{\mathrm{Att}}}

\usepackage{xcolor}

\title{Streaming Transformer Transducer Based Speech Recognition Using Non-Causal Convolution}
\name{\begin{tabular}{c}Yangyang Shi$^{*}$, Chunyang Wu$^{*}$, Dilin Wang$^{*}$, Alex Xiao, Jay Mahadeokar,  Xiaohui Zhang,\\
 Chunxi Liu, Ke Li, Yuan Shangguan, Varun Nagaraja, Ozlem Kalinli, Mike Seltzer \thanks{$\star$ Equal contribution.}\end{tabular}}
\address{Facebook AI}
\begin{document}
\ninept
\maketitle
\begin{abstract}
This paper improves the streaming transformer transducer for speech recognition by using non-causal convolution. Many works apply the causal convolution to improve streaming transformer ignoring the lookahead context. We propose to use non-causal convolution to process the center block and lookahead context separately. This method leverages the lookahead context in convolution and maintains similar training and decoding efficiency. Given the similar latency, using the non-causal convolution with lookahead context gives better accuracy than causal convolution, especially for open-domain dictation scenarios. 
Besides, this paper applies
talking-head attention and 
a novel history context compression scheme to further improve the performance. The talking-head attention improves the multi-head self-attention by transferring information among different heads. The history context compression method introduces more extended history context compactly.
On our in-house data,
the proposed methods improve a small Emformer baseline with lookahead context by relative WERR 5.1\%, 14.5\%, 8.4\% on open-domain dictation, assistant general scenarios, and assistant calling scenarios, respectively. 
\end{abstract}

\vspace{-8pt}
\section{Introduction}
\vspace{-6pt}
\label{sec:format}


Nowadays, sequence transducer networks~\cite{Graves2012,he2019rnnt} are widely used for streaming automatic speech recognition due to their superior performance and compactness. A sequence transducer model has an encoder to capture the context information from acoustic signals, a predictor to model the grammar, syntactic, and semantic information, and a joiner to combine the two parts. The work~\cite{zhang_2020,Yeh_2019} showed replacing the LSTM encoder with the self-attention-based transformer~\cite{vaswani2017} yielded the state-of-the-art of accuracy on public benchmark datasets, which is consistent to the trend in applying transformer in various scenarios for automatic speech recognition~\cite{dong2018speech,karita2019comparative,sperber2018self,zhou2018syllable,chengyi_wang_2020,frank_zhang_2020,povey2018time,yongqiang_2019_icassp,emformer,conformer,Chunyang_2020_interspeech,cfyeh_asru_2020}. 

A wide range of methods that have been proposed to improve the transformer model.
One popular variant of transformer models for speech recognition tasks is Conformer~\cite{conformer}, which adds the depth separable convolutions and macaron network structure~\cite{macaron_2019} into the transformer. The work~\cite{xie_2021_streaming_transformer,jiahui_2021_fastemit} simplified the depth-wise convolution in conformer to causal convolution to support streaming scenarios.
In~\cite{cfyeh_asru_2020}, the non-causal convolution is used to support streaming speech recognition. However, sequential block processing is required to avoid training and decoding inconsistency. The sequential block processing segments the input sequences into multiple blocks. And the model is sequentially trained on each block. The sequential block process is slow in training for low latency scenarios and incapable of dealing with large-scale dataset.

The multi-head self-attention~\cite{vaswani2017} in transformer uses different heads that conduct the attention computation separately. The attention outputs are concatenated at the end. The work~\cite{talking_heads_attention} proposed a talking-heads attention method to break the separation among different heads by inserting
two additional learnable lightweight linear projections transferring information across these heads.

The Emformer~\cite{emformer} and the augmented memory transformer~\cite{Chunyang_2020_interspeech} support streaming speech recognition by using block processing where a whole utterance is segmented into multiple blocks. The self-attention performs the computation on the current block and its surrounding left context and lookahead context. An augmented memory scheme is proposed to store the information from the previous blocks, which explicitly introduces compact long-form context while maintains limited computation and runtime memory consumption in inference. The attention output from the mean of the current block is used as a memory slot for future blocks.

In this work, we advance the Emformer~\cite{emformer} model from the following aspects. First, we leverage a similar architecture as conformer but use non-causal convolution to support streaming. In comparison with~\cite{cfyeh_asru_2020}, this work enables parallel block processing with the non-causal convolution, achieving a similar training speed as the baseline Emformer model. Second, the attention is replaced with the talking-heads attention scheme.
Third, we further simplify the augmented memory extraction process similar to~\cite{compress_transformer}, referred to as context compression: rather than using the self-attention output from the mean of each block, we directly use the linear interpolation of each block as memory. On a large-scale speech recognition task, we evaluated this novel variant of the streaming transformer.

The rest of this paper is organized as follows. In Section~2, we present the methods to advance the Emformer model. Section~3 demonstrates and analyzes the experimental results, followed by a conclusion in Section~4.



\vspace{-8pt}
\section{Methods to advance Emformer}
\vspace{-6pt}
\label{sec:format}

\begin{figure*}[htbp]
\begin{subfigure}{.5\linewidth}
\centering
\includegraphics[width=.8\linewidth]{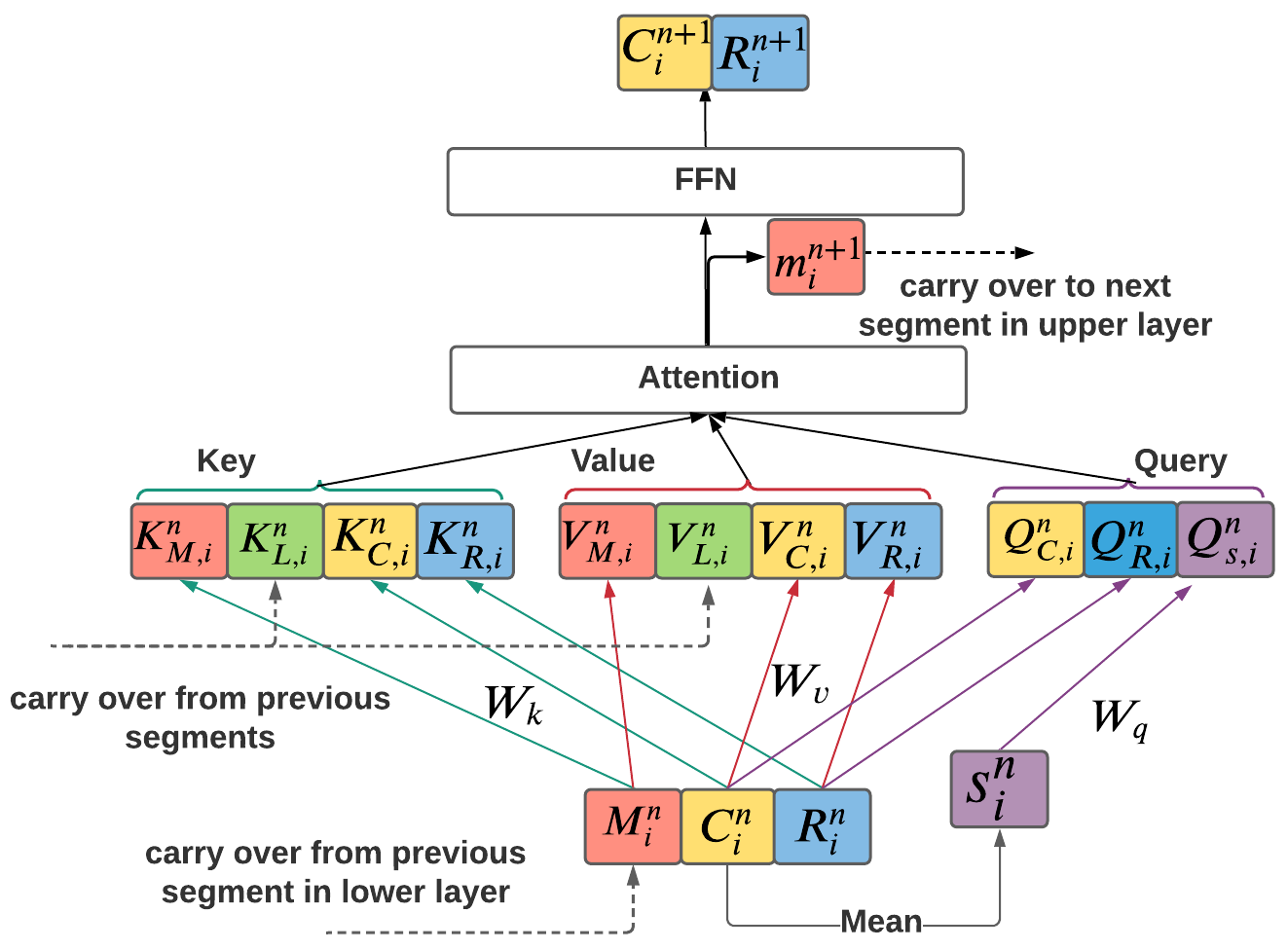}
\caption{Emformer}
\label{fig:emformer}
\end{subfigure}%
\begin{subfigure}{.5\linewidth}
\centering
\includegraphics[width=.95\linewidth]{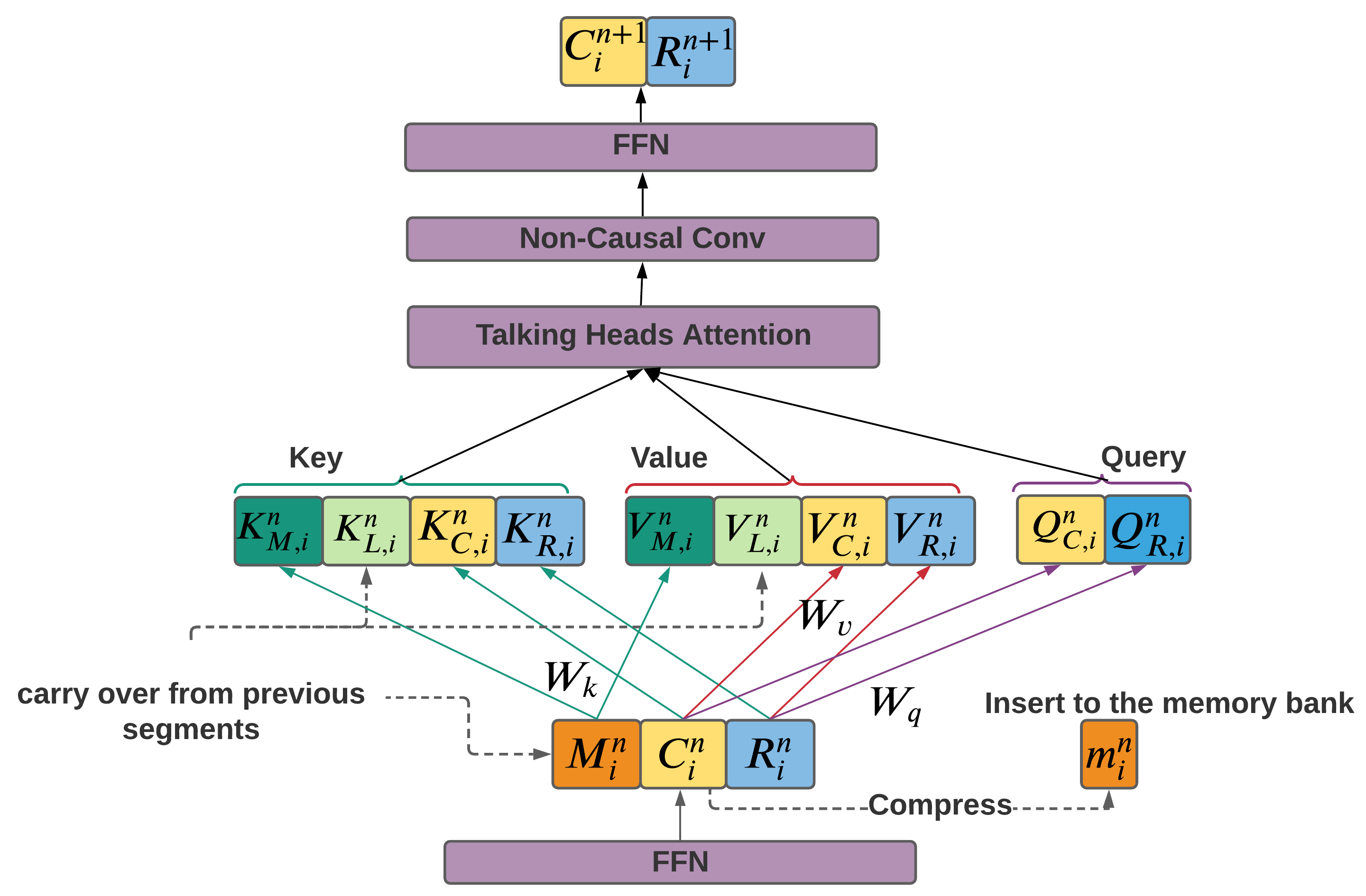}
\caption{Advanced Emformer}
\label{fig:advanced_emformer}
\end{subfigure}
\vspace{-6pt}
\caption{Advance Emformer with non-causal convolution, talking-heads attention and simplified compact augmented memory}
\vspace{-10pt}
\label{fig:advancement}
\end{figure*}

Fig.~(\ref{fig:emformer}) illustrates forward logic in one Emformer\cite{emformer} layer. To support streaming speech recognition, Emformer applies the parallel block processing to segment a input sequence into multiple non-overlapping blocks  $\mathbf{C}_1^n, \cdots, \mathbf{C}_{i-1}^n$, where $i$ denotes the index of current block, and  $n$ denotes the layer's index. In order to reduce boundary effect where the most right vector in $\mathbf{C}_i^n$ has no lookahead context information,  a right contextual block  $\mathbf{R}_i^n$, is concatenated with $\mathbf C_i^n$ to form a contextual block $\mathbf{X}_i^n = [\mathbf {C}_i^n, \mathbf R_i^n]$.
At the $i$-th block, the $n$-th Emformer layer takes $\mathbf X_i^n$ and a bank of memory vector $\mathbf{M}_i^n$ as the input, and produces $\mathbf X_{i}^{n+1} = [\mathbf C_{i}^{n+1}, \mathbf R_{i}^{n+1}]$ and $\mathbf {m}_i^n$ as the output, whereas $\mathbf X_{i}^{n+1}$ is fed to the next layer and $\boldsymbol m_i^n$ is inserted into the memory bank to generate $\mathbf M_{i+1}^{n+1}$ and carried over to the next block and next layer.

The modified attention mechanism in emformer attends to the memory bank and yields a new memory slot at each block: 
\begin{align}
 [\hat{\mathbf C}_i^n, \hat{\mathbf R}_i^n] &= \mathrm{LayerNorm}([\mathbf C_i^n, \mathbf R_i^n]), \label{input} \\
\mathbf{K}_i^n&=\left[\mathbf{K}_{L,i}^n, \mathbf{W}_{\rm k}[\mathbf{M}_i^n,   \hat{\mathbf{C}_i^n}, \hat{\mathbf{R}_i^n}]\right], \\
\mathbf{V}_i^n&=\left[\mathbf{V}_{L,i}^n, \mathbf{W}_{\rm v} [\mathbf{M}_i^n,  \hat{\mathbf{C}_i^n}, \hat{\mathbf{R}_i^n}]\right], \\
\mathbf Z_{\mathrm{C}, i}^n &= \mathrm{Attn}(\mathbf{W}_{\rm q}\hat{\mathbf C}_i^n, \mathbf{K}_i^n, \mathbf{V}_i^n) + \mathbf C_i^n,\\
\mathbf Z_{\mathrm{R}, i}^n &= \mathrm{Attn}(\mathbf{W}_{\rm q}\hat{\mathbf R}_i^n, \mathbf{K}_i^n, \mathbf{V}_i^n) + \mathbf R_i^n,\\
\mathbf{M}_i^n &= [\mathbf{m}_{i-U}^{n-1}, \cdots, \mathbf{m}_{i-1}^{n-1}], \label{eq:M0} \\
\mathbf m_i^n &= \mathrm{Attn}(\mathbf s_i^n, \mathbf{K}_i^n, \mathbf{V}_i^n), \label{eq:M1} \\
\mathbf s_i^n &= \mathrm{Mean}({\mathbf C}_i^n), \label{eq:M2}
\end{align}
where $\mathbf{K}_{L,i}^n$ and $\mathbf{V}_{L,i}^n$ are the \emph{key} and \emph{value} copies from previous blocks. $\mathbf Z_{\mathrm{C}, i}^n$ and $\mathbf Z_{\mathrm{R}, i}^n$ are the attention output for $\mathbf C_i^n$ and  $\mathbf R_i^n$ respectively; $\mathbf{s}_i^n$ is the mean of center block $\mathbf{C}_i^n$;  $\mathrm{Attn}(\mathbf q, \mathbf k, \mathbf v)$ is the attention operation defined in~\cite{vaswani2017} with $\mathbf q$ , $\mathbf k$ and  $\mathbf v$ being the query, key and value, respectively.
$U$ specifies the number of slots in augmented memory; the most recent slots are used.

$\mathbf Z_{\mathrm{C}, i}^n$ and $\mathbf Z_{\mathrm{R}, i}^n$ are passed to a point-wise feed-forward network (FFN) with layer normalization and residual connection to generate the output of this Emformer layer, i.e., 
\begin{align}
\hat{\mathbf X}_{i}^{n+1} &= \mathrm{FFN}(\mathrm{LayerNorm}([\mathbf Z_{\mathrm{C}, i}^n, \mathbf Z_{\mathrm{R}, i}^n])), \label{eqn:ffn}\\
\mathbf X_i^{n+1} &= \mathrm{LayerNorm}(\hat{\mathbf X}_{i}^{n+1} + [\mathbf Z_{\mathrm{C}, i}^n, \mathbf Z_{\mathrm{R}, i}^n]) \label{eqn:last_ln}
\end{align},
where FNN is a two-layer feed-forward network with ReLU.

\vspace{-6pt}
\subsection{Streaming Non-causal Convolution}
Fig.~(\ref{fig:advanced_emformer}) illustrates the improvements applied to advance the Emformer. Different from Eq.~(\ref{input}), the input to attention goes through one step of FFN in macaron structure:
\begin{align}
[\hat{\mathbf C}_i^n, \hat{\mathbf R}_i^n] &= \mathrm{LayerNorm}\left(\frac{1}{2}\mathrm{FFN}(\mathbf X_i^n) + \mathbf X_i^n \right).
\end{align}
Different from Eq.~(\ref{eqn:ffn}-\ref{eqn:last_ln}), the second FFN in macaron gets the input from the convolution layer.
\begin{align}
\hat{\mathbf X}_{i}^{n+1} &= \mathrm{Conv}(\mathrm{LayerNorm}([\mathbf Z_{\mathrm{C}, i}^n, \mathbf Z_{\mathrm{R}, i}^n])),\\
\mathbf X_i^{n+1} &= \mathrm{LayerNorm}\left(\hat{\mathbf X}_{i}^{n+1} + \frac{1}{2}\mathrm{FFN}(\hat{\mathbf X}_{i}^{n+1})\right).
\end{align}
The convolution layer in Fig.~(\ref{fig:advanced_emformer}) has a similar structure as \cite{conformer}, except the layer norm is used right after depth-wise convolution rather than the batch norm. In our experiments, the layer norm gives better performance than the batch norm.

The work~\cite{Chunyang_2020_interspeech,cfyeh_asru_2020} uses sequential block processing where the training and streaming decoding do the forward logic in the same way. The self-attention and convolution receptive field is limited by the block size and surrounding context size. It is trivial to use a non-causal convolution operation in this way. However, the sequential block processing is slow in training as it doesn't utilize GPU parallel computation capacity. For low latency situations where the block size is tiny, sequential block processing is not practical to use.

To use the lookahead context in streaming speech recognition, Emformer~\cite{emformer} uses the right-context-hard-copy methods in training. The right-context-hard-copy method copies and concatenates each block ${\mathrm{C}}_i^n$'s lookahead context ${\mathrm{R}}_i^n$. Then it puts the concatenated lookahead context at the beginning of the input sequence. The right-context-hard-copy method is essential to avoid the lookahead context leaking issue in training, where the higher transformer layer has a larger lookahead context than the bottom layer when multiple transformer layers are stacking on top of the other. 

\begin{figure}[h!]
\vspace{-8pt}
   \begin{center}
    \includegraphics[width=2in]{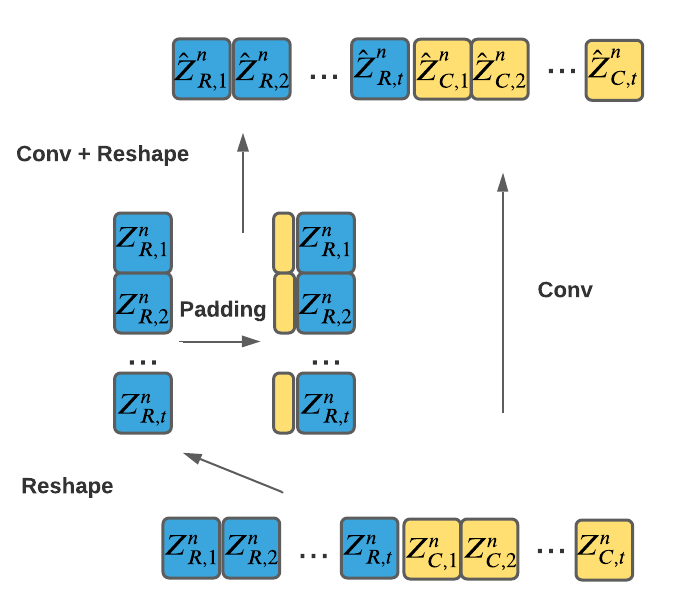}
    \end{center}
    \vspace{-16pt}
    \caption{Training procedure of non-causal convolution in parallel block processing. The same depth-wise convolution does the forward separately for lookahead context (blue) and the center block (yellow). The lookahead context is padded by feature vectors from its corresponding block.}
    \label{fig:noncausal_conv}
    \vspace{-12pt}
\end{figure}

Fig.~\ref{fig:noncausal_conv} shows the forward logic of using non-causal convolution operation in Emformer. The output from the attention operation $[\mathbf Z_{\mathrm{R}, 1}^n\dots\mathbf Z_{\mathrm{R}, t}^n, \mathbf Z_{\mathrm{C}, 1}^n\dots\mathbf Z_{\mathrm{C}, t}^n]$ is first splitted into two parts: right context $[\mathbf Z_{\mathrm{R}, 1}^n\dots\mathbf Z_{\mathrm{R}, t}^n]$ and center block $[\mathbf Z_{\mathrm{C}, 1}^n\dots\mathbf Z_{\mathrm{C}, t}^n]$. Then the same depth-wise convolution is applied to both parts. For the center block part, it is straightforward to directly apply the convolution operation as shown in Eq.~(\ref{eqn:center_block}). The right context part needs to go through reshape, padding, convolution operation and finally be reshaped to its original shape. In padding operation, each right context block is padded with its corresponding block. 
\begin{align}
[\hat{\mathbf Z}_{\mathrm{C}, 1}^n\dots\hat{\mathbf Z}_{\mathrm{C}, t}^n] &= \mathrm{Conv}([\mathbf Z_{\mathrm{C}, 1}^n\dots\mathbf Z_{\mathrm{C}, t}^n])\label{eqn:center_block} \\
\mathbf P_{\mathrm{R}, i}^n &= \mathbf Z_{\mathrm{C}, i}^n[m - k + 1: m] \label{padding} \\
[\hat{\mathbf Z}_{\mathrm{R}, 1}^n\dots\hat{\mathbf Z}_{\mathrm{R}, t}^n] &= \mathrm{Conv}([\mathbf P_{\mathrm{R}, 1}^n,\mathbf Z_{\mathrm{R}, 1}^n]\dots[\mathbf P_{\mathrm{R}, t}^n,\mathbf Z_{\mathrm{R}, t}^n])\label{eqn:right_block}
\end{align}
where $k$ is the kernel size used in depth-wise convolution. The padding $\mathbf P_{\mathrm{R}, i}^n$ in Eq.~(\ref{padding}) is the ending $k-1$ feature vectors from center block $\mathbf Z_{\mathrm{C}, i}^n$.

\newcommand{\Q}[0]{\mathbf Q}
\newcommand{\K}[0]{\mathbf K}
\newcommand{\V}[0]{\mathbf V}
\newcommand{\A}[0]{\mathbf A}
\newcommand{\W}[0]{\mathbf W}

\vspace{-9pt}
\subsection{Talking-heads Attention}
Self-attention forms the foundation of transformers. 
Assume a set of tokens $\V\in\R^{f\times d}$ that is packed into a matrix form and consider $\K \in \R^{f\times d} $ and $\V \in \R^{f\times d}$ its corresponding keys and queries, respectively. Here $f$ denotes the length of tokens and $d$ is the dimension of each token. 
Self-attention aggregates information across different tokens and transforms $V$ as follows, 
\begin{align*}
    \mathrm{Attn}(\Q, \K, \V) = \softmax\left(\Q \K^\top /\sqrt{d}\right) \V.
\end{align*}
Multi-heads attention assembles multiple standard self-attention blocks for better representation learning,
\begin{align*}
    \mathrm{MultiHeadAttn}(\Q, \K, \V) =  \mathrm{Concat}\bigg(\bigg\{\att(\Q_i, \K_i, \V_i) \bigg\}_i^h\bigg),
\end{align*}
where $h$ denotes the number of heads and $\Q_i$, $\K_i$ and $\V_i$ represents the queries, keys and values from different heads, respectively. 

One potential drawback of multiple-attention is that different heads are trained independently without coordination. Talking-heads attention~\cite{talking_heads_attention} improves on multi-heads attention by allowing information fusion among different attention heads.  
Assume $\softmax(\A)$ the attention weights learned by different heads in multi-head attention ($\A \in \R^{f\times f\times h}$).
Talking-heads attention introduces two additional linear layers immediately before and after the softmax and computes the new self-attention weights as follows,
\begin{align}
\softmax(\A * \W_l) * \W_r.
\end{align}
Here $ \W_l \in \R^{h\times h} $ and $ \W_r \in \R^{h\times h}$ are trainable parameters, and $\softmax$ is applied on the second dimension.
In practice, the two linear projections introduced by talking-heads attention are computationally efficient as the number of heads $h$ used is often small. 

\vspace{-6pt}
\subsection{Context Compression}

The augmented memory is designed to introduce long-form information into the attention. As shown in Eq.~(\ref{eq:M0}-\ref{eq:M1}), the information is introduced via the queries of the previous segments in the previous layer.
This inter-layer strategy gets rid of the auto-regression property if it is on the same layer, preventing inefficient block processing in training.
However, one potential issue of this design is the representation mismatch between successive layers.
In the attention operation, the augmented memory slots from the previous layer and the frames from the current layer are equally treated in key and value, which depends on the similar representations on the two layers.
Otherwise, 
long-form information can be misleadingly introduced.

To address the potential mismatch between memory slots and frames,
we put forward the context compression strategy in this paper.
The context compression directly introduces compact memory to the key and value in the attention, not to the query. It is formalized as follows,
\begin{align}
\mathbf m_i^n &= \mathrm{Compress}({\mathbf C}_i^n), \label{eq:cm} \\
\mathbf{M}_i^n &= [\mathbf{m}_{i-U}^{n}, \cdots, \mathbf{m}_{i-1-O}^{n}]  \label{eq:cmg}
\end{align}
where the $\mathrm{Compress}$ operation stands for a function that can compress the segment into one single vector, e.g., linear interpolation or average pooling; this work chooses the linear interpolation. Contrasting to Eq.~(\ref{eq:cmg}), an offset term $O$ is introduced in Eq.~(\ref{eq:M0}), which is intended to prevent the overlap between the short-form left context and this long-form compressed context. For instance, on a model with a segment size of 4 and a left context of 8, we set an offset of 2 to skip the interval covered by the left context.
According to Eq.~(\ref{eq:cm}), the context compression operates the input $C$ of each layer, preventing the auto-regression between successive segments;
thus the whole sequence can be trained in parallel, thoroughly taking advantage of the graphics computing resource.

\vspace{-3pt}
\section{Experiments}
\label{sec:pagestyle}
\vspace{-3pt}
\subsection{Data}
\vspace{-6pt}
Our training data is a large-scale speech recognition dataset composed of two scenarios. The \textit{assistant} scenario consists of three parts. One is 13K hours of recordings collected from third-party vendors via crowd-sourced volunteers responding to artificial prompts with mobile devices. The content varies from voice assistant commands to a simulation of conversations between people. The second is 1.3K hours of voice commands from production. The last is 4K hours of speech for calling names and phone numbers generated by an in-house TTS model. The \textit{open domain} dictation has 18K hours of human transcribed data from video and 2M hours of unlabeled videos transcribed by a high-quality in-house model. The data was augmented with various distortion methods: speed perturbation~\cite{ko2015audio}, simulated reverberation SpecAugment \cite{park2019specaugment}, and randomly sampled additive background noise extracted from videos.

In evaluation, we use \textit{assi}, \textit{call} and \textit{dict} dataset. The \textit{assi} and \textit{call} are 13.6K manually transcribed utterances from in-house volunteer employees, and each utterance starts with a wake word. The \textit{dict} is 8 hours open domain dictation from crowd-sourced workers recorded via mobile devices.

\vspace{-6pt}
\subsection{Experiment Setting}

The input features are 80-dim log Mel filter bank features at a 10ms frame rate; The network's input is 
a 640-dim superframe consists of 8 consistent frames with a downsampling factor of 8 to 80ms frame rate. This paper explored models with 32M parameters and 73M parameters. In the 32M parameter baseline model, a projection layer maps the superframe to a 320-dim vector. The encoder consists of 21 Emformer layers. Each layer uses four heads for self-attention, and its FFN-block dimension is 1280. The predictor consists of a 256-dim embedding layer with 4096 sentence pieces~\cite{kudo2018sentencepiece}, 1 LSTM layer with 512 hidden nodes, and a linear projection layer with 1024 output nodes. 
The baseline with 32M parameters uses a left context of 640ms (10 slots) in the left context.
For the block size and right context, two settings are investigated.
One is the block size of 320ms (4 slots) and right context of 80ms (1 slot);
the other is a block size of 400ms (5 slots) and a right context of 0 (0 slots). In the 73M parameter model, the superframe is mapped to a 512-dim vector. The encoder has 20 layers of Emformer.
Each layer has an 8-head self-attention and a 2048-dim FFN block.
Its predictor has the same layer configuration as the 32M baseline, but the number of LSTM layers is 3. The left context is set to 2.4s, i.e., 30 slots in the left context. In training, on the 73M parameter model, \emph{SpecAugment}~\cite{park2019specaugment} without time warping, and dropout 0.1 are used.  We found that the 32M parameter models are underfitting a large amount of training data. The best performance is obtained by not using either scheme.

For our proposed models, we first investigate the non-causal convolution.
A kernel size seven is used for depth-wise convolution operations. In the 32M parameter model,  the superframe is projected to a 256-dim vector.
In the 73M parameter model, the superframe is projected to a 384-dim vector. It consists of 20 layers containing an 8-head self-attention and a 1456-dim FFN block in each layer. It consists of 18 layers containing a 4-head self-attention and a 1024-dim FFN block in each layer. Other settings are the same as the baselines. The block size and right context are fixed as 320ms and 80ms, respectively. For the context compression scheme, we use a regular left context of 8 slots, implying 640ms. The compressed left context is set to 2 slots, implying 640ms; also, it uses an offset of 2, $O$ in Eq.~(\ref{eq:cmg}), to skip the same interval of the 640ms regular left context. In total, ten slots are used but implying a history of 1280ms.

In all the experiments, alignment restrict RNNT~\cite{jay_2020_arrnnt} is used. 
The training of all the models uses 32 Nvidia V100 GPUs. We evaluate the models by word error rate (WER) for accuracy and the real-time factors (RTFs) and speech engine perceived latency (SPL) for latency. The SPL measures the time the speech engine gets the last word from user utterance to the speech engine transcribes the last word and gets the endpoint signals.

\vspace{-6pt}
\subsection{Improvement from Non-causal Convolution}
Table.~\ref{tab:32m_noncausal} gives the WER, RTF, and SPL results for models with 32M and 73M parameters. The results show that by keeping the overall context size (the sum of block size and lookahead context size) the same, using lookahead context gives WER improvement over not using it, especially for open domain dictation scenarios. We also observe that convolution and macaron structure improves the baseline using the same context configuration. Table.~\ref{tab:32m_noncausal} also show that the direct application of causal convolution with 400ms block size does not improve the baseline model which leverages 320ms block size with 80ms lookahead context for both 32M and 73M models. 

Using lookahead context adds more computation for encoders in transducer model, as the forward logic has duplicated computation for the lookahead context. For the 73M model, using right side context 80ms shows $10\%$ relative RTF increase. However, lookahead context provides more accurate ASR results and yields slightly better speech perceived latency (SPL).

\vspace{-3pt}
\begin{table}[htbp]
    \centering
    \caption{WER, RTF and SPL impact from non-causal convolution and lookahead context.Column `\#p' gives the number of parameters in each model. Column `w/C' denotes whether the convolution is applied or not. `C' and `R' represents the block size and lookahead context size. The unit for `C', `R' and `SPL' is millisecond.}
    \begin{tabular}{|cccc|ccc|cc|}
    \hline
   \#p & w/c &C &R&\textit{dict} & \textit{ass} & \textit{call} & SPL &RTF\\
   \hline\hline
    \multirow{2}{*}{73M}& \multirow{2}{*}{N} & 400 & 0 & 16.78 & 4.18 & 6.19 &606 &0.27\\
    & & 320 & 80 & 15.49 & 3.98 & 5.81 &599 &0.30\\
    \hline
    \multirow{2}{*}{73M} & \multirow{2}{*}{Y} & 400 & 0 &16.11  &4.05  &5.94  &615 &0.27\\
    & & 320 & 80 & 14.67 & 3.65 & 5.85 &595 &0.30\\
   \hline\hline
    \multirow{2}{*}{32M} &\multirow{2}{*}{N} & 400 & 0 & 18.31& 5.17 & 6.57 &635 &0.21\\
     & & 320 & 80 & 17.09 & 5.05 & 6.68 &588 &0.22\\
    \hline
    \multirow{2}{*}{32M}& \multirow{2}{*}{Y} & 400 & 0 &17.70  &4.78  &6.76  &626 &0.22\\
    & & 320 & 80 & 17.02 & 4.66 & 6.53 &605 &0.22\\
    \hline 
    \end{tabular}
    \label{tab:32m_noncausal}
\end{table}

\subsection{Improvement from Talking-heads Attention and Context Compression}
Table.~\ref{tab:ablation} shows the impact of applying talking-heads attention and context compression on top of Emformer with non-causal convolutions. For the model with 32M parameters, the talking-heads attention generates $4.6\%$, $3.8\%$, and $2.8\%$ relative WER reductions on open-domain dictation, assistant general queries, and assistant calling queries, respectively.  Using two slots of context compression outperforms the model with only regular left context. Combining non-casual convolution, talking-heads, and context compression in the 32M model improves the WER by 5.1\%, 14.5\%, 8.4\% relatively on open-domain dictation, assistant general, and assistant calling test sets, while maintaining similar SPL and RTF as the Emformer baseline. For the model with 73M parameters, talking-heads attention and context compression obtain on par WER as the Emformer with non-causal convolution. Note the 73M parameters baseline uses 30 slots of left context, while context compression uses 8 slots of left context and 2 slots of memory which slightly improves the RTF and SPL. However, the 73M model already has a much stronger model capacity than the 32M model. The lightweight optimizations of talking-heads attention and context compression do not generate obvious improvement.

\begin{table}[htbp]
\vspace{-3pt}
    \centering
    \caption{WER and RTF and SPL impact from the context compression. Column `L' stands for the length of left context. Column `CL' stands for the length of compressed left context. The unit of both columns is slot: for `L', 1 slot means 80ms; for `CL', 1 slot implies 320ms, i.e. block size. 
    }
    \begin{tabular}{|l|ccc|cc|}  
    \hline
     Model &\textit{dict} & \textit{ass} & \textit{call} & SPL & RTF\\ 
    \hline
     \hline
    73M Emformer (baseline) & 15.49 & 3.98 & 5.81 &599 &0.30 \\
    \hline
    + Non-causal & 14.67 & 3.65 & 5.85 &595 &0.30 \\
    + Talk heads & 14.69  & 3.64 & 5.79 &574 & 0.30 \\
    + Context Compression& 14.69 & 3.66 & 5.77 & 554 & 0.29 \\
    \hline
    32M Emformer (baseline) & 17.09 & 5.05 & 6.68 &588 &0.22 \\
    \hline
    + Non-causal &  17.02 & 4.66 & 6.53 &605 &0.22 \\
    + Talk heads & 16.25 &  4.48 &  6.35 &620  &0.23 \\
    + Context Compression& 16.22 & 4.32 & 6.12 & 589 & 0.24 \\
    \hline   
    \end{tabular}
    \label{tab:ablation}
\end{table}

\vspace{-18pt}
\section{Conclusions}
\vspace{-6pt}
In this work, we proposed to use non-causal convolution, talking heads attention, and context compression to improve the streaming transformer transducer for speech recognition. This work managed to apply non-causal convolution with lookahead context in streaming transformer by separating the forward logic for the center block and lookahead context. The talking-heads attention coordinates the training of different heads in self-attention. 
The context compression keeps the representation contained in the long-form and short-form history similar, providing a compact way of introducing long-form information. The experiments on 32M parameter and 73M parameter models show that 
the proposed model outperforms the Emformer baseline on open-domain dictation, assistant general, and assistant calling scenarios while maintaining comparable RTF and latency.

\bibliographystyle{IEEEbib}
\bibliography{refs}

\begin{thebibliography}{10}

\bibitem{Graves2012}
Alex Graves,
\newblock ``{Sequence Transduction with Recurrent Neural Networks},''
\newblock {\em arXiv preprint arXiv:1211.3711}, 2012.

\bibitem{he2019rnnt}
Y~He, T~N Sainath, R~Prabhavalkar, et~al.,
\newblock ``{Streaming End-to-end Speech Recognition for Mobile Devices},''
\newblock in {\em Proc. ICASSP}, 2019.

\bibitem{zhang_2020}
Qian Zhang, Han Lu, Hasim Sak, Anshuman Tripathi, Erik McDermott, Stephen Koo,
  and Shankar Kumar,
\newblock ``{Transformer Transducer: A Streamable Speech Recognition Model with
  Transformer Encoders and RNN-T Loss},''
\newblock in {\em Proc. ICASSP}, 2020.

\bibitem{Yeh_2019}
Ching-Feng Yeh, Jay Mahadeokar, Kaustubh Kalgaonkar, and Others,
\newblock ``{Transformer-Transducer: End-to-End Speech Recognition with
  Self-Attention},''
\newblock {\em arXiv preprint arXiv:11910.12977}, 2019.

\bibitem{vaswani2017}
Ashish Vaswani, Noam Shazeer, Niki Parmar, Jakob Uszkoreit, Llion Jones,
  Aidan~N. Gomez, {\L}ukasz Kaiser, and Illia Polosukhin,
\newblock ``{Attention is all you need},''
\newblock in {\em Proc. NIPS}, 2017.

\bibitem{dong2018speech}
L~Dong, S~Xu, and B~Xu,
\newblock ``{Speech-transformer: a no-recurrence sequence-to-sequence model for
  speech recognition},''
\newblock in {\em Proc. ICASSP}, 2018.

\bibitem{karita2019comparative}
S~Karita, N~Chen, T~Hayashi, and Others,
\newblock ``{A Comparative Study on Transformer vs RNN in Speech
  Applications},''
\newblock {\em arXiv preprint arXiv:1909.06317}, 2019.

\bibitem{sperber2018self}
M~Sperber, J~Niehues, G~Neubig, et~al.,
\newblock ``{Self-attentional acoustic models},''
\newblock {\em arXiv preprint arXiv:1803.09519}, 2018.

\bibitem{zhou2018syllable}
S~Zhou, L~Dong, S~Xu, and B~Xu,
\newblock ``{Syllable-based sequence-to-sequence speech recognition with the
  transformer in mandarin Chinese},''
\newblock {\em arXiv preprint arXiv:1804.10752}, 2018.

\bibitem{chengyi_wang_2020}
Chengyi Wang, Yu~Wu, Shujie Liu, Jinyu Li, Liang Lu, Guoli Ye, and Ming Zhou,
\newblock ``{Low Latency End-to-End Streaming Speech Recognition with a Scout
  Network},''
\newblock {\em arXiv preprint arXiv:12003.10369}, 2020.

\bibitem{frank_zhang_2020}
Frank Zhang, Yongqiang Wang, Xiaohui Zhang, Chunxi Liu, Yatharth Saraf, and
  Geoffrey Zweig,
\newblock ``{Fast, Simpler and More Accurate Hybrid ASR Systems Using
  Wordpieces},''
\newblock {\em InterSpeech}, 2020.

\bibitem{povey2018time}
D~Povey, Hossein Hadian, P~Ghahremani, and Others,
\newblock ``{A time-restricted self-attention layer for asr},''
\newblock in {\em Proc. ICASSP}, 2018.

\bibitem{yongqiang_2019_icassp}
Yongqiang Wang, Abdelrahman Mohamed, Due Le, Chunxi Liu, Alex Xiao, Jay
  Mahadeokar, Hongzhao Huang, Andros Tjandra, Xiaohui Zhang, Frank Zhang,
  Christian Fuegen, Geoffrey Zweig, and Michael~L. Seltzer,
\newblock ``{Transformer-Based Acoustic Modeling for Hybrid Speech
  Recognition},''
\newblock in {\em Proc. ICASSP}, 2019.

\bibitem{emformer}
Yangyang Shi, Yongqiang Wang, Chunyang Wu, Ching-Feng Yeh, and Others,
\newblock ``{Emformer: Efficient Memory Transformer Based Acoustic Model For
  Low Latency Streaming Speech Recognition},''
\newblock in {\em Proc. ICASSP}, 2021.

\bibitem{conformer}
Anmol Gulati, James Qin, Chung~Cheng Chiu, et~al.,
\newblock ``{Conformer: Convolution-augmented transformer for speech
  recognition},''
\newblock in {\em Proc. INTERSPEECH}, 2020.

\bibitem{Chunyang_2020_interspeech}
Chunyang Wu, Yang Shi, Yongqiang Wang, and Ching-Feng Yeh,
\newblock ``{Streaming Transformer-based Acoustic Modeling Using Self-attention
  with Augmented Memory},''
\newblock in {\em Proc. INTERSPEECH}, 2020.

\bibitem{cfyeh_asru_2020}
Ching~Feng Yeh, Yongqiang Wang, Yangyang Shi, et~al.,
\newblock ``{Streaming attention-based models with augmented memory for
  end-to-end speech recognition},''
\newblock in {\em Proc. SLT}, 2020.

\bibitem{macaron_2019}
Yiping Lu, Zhuohan Li, Di~He, Zhiqing Sun, Bin Dong, Tao Qin, Liwei Wang, and
  Tie-Yan Liu,
\newblock ``{Understanding and Improving Transformer From a Multi-Particle
  Dynamic System Point of View},''
\newblock {\em arXiv preprint arXiv:1906.02762}, 2019.

\bibitem{xie_2021_streaming_transformer}
Xie Chen, Yu~Wu, Zhenghao Wang, Shujie Liu, and Jinyu Li,
\newblock ``{Developing Real-Time Streaming Transformer Transducer for Speech
  Recognition on Large-Scale Dataset},''
\newblock in {\em Proc. ICASSP}, 2021, pp. 5904--5908.

\bibitem{jiahui_2021_fastemit}
Jiahui Yu, Chung-Cheng Chiu, Bo~Li, and Others,
\newblock ``{Fastemit: Low-Latency Streaming Asr With Sequence-Level Emission
  Regularization},''
\newblock in {\em Proc. ICASSP}, 2021, vol.~53.

\bibitem{talking_heads_attention}
Noam Shazeer, Zhenzhong Lan, Youlong Cheng, Nan Ding, and Le~Hou,
\newblock ``{Talking-Heads Attention},''
\newblock {\em arXiv preprint arXiv:2003.02436}, 2020.

\bibitem{compress_transformer}
Jack~W. Rae, Anna Potapenko, Siddhant~M. Jayakumar, and Timothy~P. Lillicrap,
\newblock ``{Compressive Transformers for Long-Range Sequence Modelling},''
\newblock {\em arXiv preprint arXiv:1911.05507}, 2019.

\bibitem{ko2015audio}
T~Ko, V~Peddinti, D~Povey, and Others,
\newblock ``{Audio augmentation for speech recognition},''
\newblock in {\em Proc. INTERSPEECH}, 2015.

\bibitem{park2019specaugment}
D~S Park, W~Chan, Y~Zhang, et~al.,
\newblock ``{Specaugment: A simple data augmentation method for automatic
  speech recognition},''
\newblock {\em arXiv preprint arXiv:1904.08779}, 2019.

\bibitem{kudo2018sentencepiece}
Taku Kudo and John Richardson,
\newblock ``{SentencePiece: A simple and language independent subword tokenizer
  and detokenizer for neural text processing},''
\newblock {\em Proc. EMNLP}, 2018.

\bibitem{jay_2020_arrnnt}
Jay Mahadeokar, Yuan Shangguan, Duc Le, and Others,
\newblock ``{Alignment restricted streaming recurrent neural network
  transducer},''
\newblock in {\em Proc. SLT}, 2021.

\end{thebibliography}

\end{document}